\documentclass[trackchanges,twocolumn]{aastex7}

\usepackage{amsmath}
\usepackage{array}

\begin{document}

\title{Identifying close-in Jupiters that arrived via disk migration:\\ Evidence of primordial alignment, preference of nearby companions and hint of runaway migration}

\author[orcid=0000-0002-0488-6297,sname="Kawai"]{Yugo Kawai}
\affiliation{Department of Multi-Disciplinary Sciences, Graduate School of Arts and Sciences, The University of Tokyo, 3-8-1 Komaba, Meguro, Tokyo 153-8902, Japan}
\email[show]{yugo6581@g.ecc.u-tokyo.ac.jp}  

\author[orcid=0000-0002-4909-5763]{Akihiko Fukui}
\affiliation{Komaba Institute for Science, The University of Tokyo, 3-8-1 Komaba, Meguro, Tokyo 153-8902, Japan}
\affiliation{Instituto de Astrof\'{i}sica de Canarias (IAC), 38205 La Laguna, Tenerife, Spain}
\email{afukui@g.ecc.u-tokyo.ac.jp}  

\author[orcid=0000-0002-7522-8195]{Noriharu Watanabe}
\affiliation{Department of Multi-Disciplinary Sciences, Graduate School of Arts and Sciences, The University of Tokyo, 3-8-1 Komaba, Meguro, Tokyo 153-8902, Japan}
\email{n-watanabe@g.ecc.u-tokyo.ac.jp} 

\author{Sho Fukazawa}
\affiliation{Department of Multi-Disciplinary Sciences, Graduate School of Arts and Sciences, The University of Tokyo, 3-8-1 Komaba, Meguro, Tokyo 153-8902, Japan}
\email{yugo6581@g.ecc.u-tokyo.ac.jp}  

\author[orcid=0000-0001-8511-2981]{Norio Narita}
\affiliation{Komaba Institute for Science, The University of Tokyo, 3-8-1 Komaba, Meguro, Tokyo 153-8902, Japan}
\affiliation{Instituto de Astrof\'{i}sica de Canarias (IAC), 38205 La Laguna, Tenerife, Spain}
\affiliation{Astrobiology Center, 2-21-1 Osawa, Mitaka, Tokyo 181-8588, Japan}
\email{yugo6581@g.ecc.u-tokyo.ac.jp}

\begin{abstract}

Two leading hypotheses for hot Jupiter migration are disk migration and high-eccentricity migration (HEM). Stellar obliquity is commonly used to distinguish them, as high obliquity often accompanies HEM. However, low obliquity does not guarantee disk migration, due to possible spin-orbit realignment or coplanar HEM. Seeking a proxy for disk migration, we investigate the idea that when the circularization timescale of a planet on circular orbit is longer than its age ($\tau_\mathrm{cir} > \tau_\mathrm{age}$), HEM would not have had sufficient time to complete, favoring disk migration. We empirically calibrate the reduced planetary tidal quality factor to be $Q_\mathrm{p}=4.9^{+3.5}_{-2.5}\times10^5$ using the eccentricity distribution of 500+ Jovian mass ($0.2M_\mathrm{J}<M_\mathrm{p}<13M_\mathrm{J}$) planets with measured masses and radii, a value consistent with solar system Jupiter. We then calculate $\tau_\mathrm{cir}$ and identify dozens of disk migration candidates ($\tau_\mathrm{cir} > \tau_\mathrm{age}, \ e < 0.1$). These planets show three notable trends. We first find a clear cutoff of obliquity at $\tau_\mathrm{cir} \sim \tau_\mathrm{age}$, suggesting the primordial alignment of protoplanetary disks. Secondly, we find that among hot Jupiters ($a<0.1$ au), nearby companions are preferentially found around disk migration candidates, suggesting that either HEM dominates hot Jupiter formation, or disk migration also disrupts nearby companions at short separations. Finally, we find a possible dearth of disk migration candidates around mass ratio $\log q \sim -3.2$, consistent with a similar dip suggested at longer orbits from microlensing. The lack of planets across different orbital distance, if true, could be interpreted as a hint of runaway migration.
\end{abstract}



\section{Introduction} 
Hot Jupiters are a peculiar class of gas giant exoplanets orbiting extremely close to their host stars, raising longstanding questions about their formation. In-situ formation, although explored extensively in some works \citep[e.g.][]{Batygin_2016}, is considered difficult due to local disk conditions, suggesting instead that these planets formed at larger orbital distances and migrated inward \citep{Rafikov_2005}. Consequently, two primary mechanisms have been proposed to explain their existence: (1) disk migration through planet-disk interactions \citep[e.g.][]{Tanaka_2002,Trilling_2002}, and (2) high-eccentricity migration (HEM) \citep[e.g.][]{Fabrycky_Tremaine_2007,Ford_Rasio_2008,Wu_Lithwick_2011}, in which gravitational perturbations from a stellar or planetary companion excite large orbital eccentricities, followed by tidal circularization near periastron.

Measurement of stellar obliquity, which is likely excited concurrently with orbital eccentricity, has been a common method of testing the two hypotheses. But its efficacy has been limited by confounding scenarios associated with both high obliquity (HEM) and low obliquity (disk migration). On one hand, high obliquity is also achieved with primordial misalignement of protoplanetary disks. For instance, gravitational interactions between host stars, protoplanetary disks, and inclined binary companions can create such misalignments \citep{Batygin_2013,Lai_2014,Spalding_2014,Zanazzi_2018}. Observationally, misaligned inner disks \citep{Marino_2015,Ansdell_2020}, coplanar but misaligned multi-planet systems \citep{Huber_2013,Hjorth_2021} and extremely large obliquity not reproducible with ordinary HEM \citep{Kawai_2024}, albeit rare, provide evidence for such scenarios.

More critically, low obliquity does not necessarily point to disk migration. This is primarily due to the fact that obliquity could damp due to tidal dissipation in the star and hot Jupiters that arrived via HEM could very well have aligned orbits as well \citep{Albrecht_2012,Rice_2022}. In other instances, high eccentricity can also be achieved without exciting obliquity in a process known as coplanar high eccentricity migration (CHEM) \citep{Li+2014,Petrovich+2015}.
 
We hence lack good proxies for disk migration, making observational tests of migration theories difficult. The BOWIE-ALIGN survey is an ongoing JWST observation campaign aimed at a comparative study of C/O ratio between hot Jupiters from disk migration and HEM (i.e. disk-free migration) \citep{Kirk+2024}. The target selection is based on obliquity, and leverages on the host stellar effective temperature to rule out targets susceptible to spin-orbit realignment. However, the possibility of CHEM still complicates the sample selection using obliquity. 

One promising proxy for disk migration is the existence of nearby planetary companions, as the violent nature of HEM will likely disrupt the orbits of such planets. The discoveries of such hot Jupiter systems with nearby companions are increasing (e.g. WASP-47: \citealt{Becker+2015} , WASP-84: \citealt{Maciejewski_2023}, WASP-132 \citealt{Hord_2022}, TOI-1408: \citealt{Korth+2024}, TOI-1130: \citealt{Korth+2023}, TOI-2000: \citealt{Sha+_2023}, TOI-2202: \citealt{Trifonov_2021}, Kepler-730 \citealt{Canas+2019}), but are still rare. It is uncertain if the scarcity of nearby companions suggest that HEM is the dominant mode of migration, as disk migration can too disrupt nearby companions in the process \citep{Wu_He_2023, He_Wu_2024}. The scarcity is nonetheless the limitation of using nearby companions as proxy for disk migration.

In this study, the proxy of interest is $\tau_\mathrm{cir}$, the circularization timescale of the planetary orbit. Because the assumption in HEM is that tidal circularization (on top of eccentricity excitation) completes within the system age, systems where $\tau_\mathrm{cir} > \tau_\mathrm{age}$ but eccentricity $e \sim 0$, is incompatible with HEM. Most uncertainty in $\tau_\mathrm{cir}$ lies in the tidal quality factor $Q_\mathrm{p}$, which encapsulates the efficiency of tidal dissipation in the planet \citep[e.g.][]{Goldreich_Soter_1966}. Since the planetary love number $k_2$ is also uncertain, we adopt the more conventional reduced tidal quality factor $Q'_\mathrm{p} =3Q_\mathrm{p}/2k_2$, and omit the dash for brevity. Once $Q_\mathrm{p}$ is constrained, $\tau_\mathrm{cir}$ to the simplest approximation of equilibrium tides can be estimated with only basic system parameters like planet mass and radius. Therefore, we can expect $\tau_\mathrm{cir}$ to be a very accessible proxy for testing migration scenarios, and its viability hinges on how well we can constrain $Q_\mathrm{p}$.

In our solar system, Jupiter's tidal quality factor is constrained to $6\times10^4<Q_\mathrm{J}<2\times10^6$ for Io to remain in Laplace resonance with Ganymede and Europa \citep{YoderPeale_1981}, assuming the resonance is not primordial. For exoplanets, \cite{Matsumura_2008} first compared $\tau_\mathrm{cir}$ of known hot Jupiters at the time to their stellar ages and individually constrained their $Q_\mathrm{p}$. Assuming that all circular orbits is the result of tidal circularization, they placed lower limits on $Q_\mathrm{p}$ of circular hot Jupiters to satisfy $\tau_\mathrm{cir} < \tau_\mathrm{age}$. Similarly, they placed upper limits on $Q_\mathrm{p}$ for eccentric systems to satisfy $\tau_\mathrm{cir} > \tau_\mathrm{age}$. As a result, they found $10^5<Q_\mathrm{p}<10^9$ in general. \cite{Jackson+2008}, using backward integration of hot Jupiters with $a<0.2\mathrm{AU}$, estimated $Q_\mathrm{p} \sim10^{6.5}$ as value that best reproduces the eccentricity distribution of Jupiters with $a>0.2\mathrm{AU}$. 

\cite{Hansen_2010} and \cite{Hansen_2012} on the other hand, derived $Q_\mathrm{p} \sim10^{7-8}$ from a forward evolution of simulated Jupiters to match the observed eccentricity and orbital period distribution of observed Jupiters. Interestingly, \citet{Hansen_2010} observed the presence of hot and warm Jupiters on circular orbits, despite having orbital periods that exceed the so-called ``circularization period'', $P_\mathrm{cir}$, —an empirical boundary that separates planets with circular orbits from those with eccentric orbits, a concept more commonly applied to stellar binaries \citep[e.g.][]{Meibom_Mathieu_2005}. They proposed that these circular Jupiters might be a result of disk migration, but also noted that selection biases in transit and RV surveys could have inflated their detectability. While \citet{Hansen_2010} did not explore the presence of such circular Jupiters beyond the circularization period in detail, the focus of our work is to identify circular planets with \( \tau_\mathrm{cir} > \tau_\mathrm{age} \), which is conceptually similar to identifying circular ones with \( P > P_\mathrm{cir} \).

\cite{Quinn+2014} (hereafter Q14) derived a range of \( 8 \times 10^5 < Q_\mathrm{p} < 3 \times 10^6 \) as the value that maximizes the difference in eccentricity distributions between dynamically young and dynamically old systems. In this context, \emph{dynamically young} refers to systems with \( \tau_\mathrm{cir} > \tau_\mathrm{age} \), where tidal circularization is expected to be incomplete within the system's lifetime, whereas \emph{dynamically old} refers to those with \( \tau_\mathrm{cir} < \tau_\mathrm{age} \), where circularization should already have completed.\footnote{The goal of Q14 was not to identify disk migration candidates, but to demonstrate that dynamically young and old systems exhibit different eccentricity distributions, suggesting tidal circularization in effect.} The approach is conceptually similar to that of \cite{Matsumura_2008}, but Q14 aims to constrain the population-wide value of \( Q_\mathrm{p} \) rather than estimate it on a per-planet basis. As emphasized in Q14, a key advantage of this method is that it remains valid regardless of the relative fraction of disk migration versus HEM; as long as tidal circularization is active in some subset of systems, an optimal value of \( Q_\mathrm{p} \) should emerge. In contrast, attempts to constrain \( Q_\mathrm{p} \) on an individual basis becomes unreliable for hot Jupiters whose circular orbits are primordial and not the result of tidal circularization. Leveraging the growing sample of transiting hot Jupiters with well-measured masses and eccentricities, we revisit the methodology of Q14 and extend it to identify candidates likely formed via disk migration.

This paper is structured as follows. In Section \ref{sec:methods}, we describe the methods including the sample selection procedure and our implementation of the algorithm used to constrain $Q_\mathrm{p}$ which follows Q14. Section \ref{sec:results} describes the results of this study, and we discuss those results in Section \ref{sec:discussion}.

\section{Methods} \label{sec:methods}
\subsection{Sample selection} \label{sec:sample_selection}
Our sample consists of Jovian mass planets with masses in the range $0.2M_\mathrm{J}<M_\mathrm{p}<13M_\mathrm{J}$, radius larger than $R_\mathrm{p}>5R_\oplus$ and an orbital period of less than 365 days. Including planets with longer orbital periods allows us to explore a wide range of $Q_\mathrm{p}$, but the choice of 365 days is arbitrary. As it turns out, the threshold between dynamically young and old planets from our experiment lie at orbital periods of around 5 to 10 days. So the inclusion of planets with orbital periods of hundreds of days are not vital, but the inclusion of planets with a few tens of days plays a role in determining $Q_\mathrm{p}$.

We also limit our samples to transiting systems with radius measurements so that $\tau_\mathrm{cir}$ can be calculated without the use of empirical mass-radius relations, which can be very degenerate for Jovian mass planets. We obtain the full list of planets matching the above criteria from the NASA exoplanet archive's Confirmed Planets Table as of May 22, 2025, retrieving the default parameters. We then manually update the mass and eccentricity constraints for over 200 planets with values from \cite{Bonomo+2017}, when those parameters are not set to the default. This left us with 490 planets with either fully constrained eccentricity or with upper limits. 

We then manually inspected the 166 planets with assumed circular orbits (i.e. $e=0$ with no error reported) in the default parameter set. Of those, eccentricity free fits were performed for 88 planets but circular solution was adopted either based on statistical model selections or due to lack of sufficient data points. Indeed, small eccentricities can be statistical in origin and not significant \citep{Lucy_Sweeney_1971}. However, we find that some planets with assumed circular orbits lie on the borderline between being dynamically young and old, and argue that in such cases, the assumption of circular orbits should be reconsidered. We therefore remain conservative and use the constrains given from the eccentricity free fits for the 88 planets, and exclude the ones on assumed circular orbits from the sample. This resulted in 578 planets with measured eccentricity. We then removed 41 systems without age constraints, resulting in 537 systems for which we used to calibrate $Q_\mathrm{p}$.  $\tau_{\mathrm{cir}}$ is calculated for all 668 planets in the sample.

\subsection{Constraining $Q_\mathrm{p}$} \label{sec:constraining_qp}

Following Q14, $Q_\mathrm{p}$ in this work is constrained leveraging on the fact that for planets starting on eccentric orbits, there should exist a circularization boundary at $\tau_{\mathrm{cir}}\sim\tau_{\mathrm{age}}$, where $\tau_{\mathrm{cir}}$ is the circularization timescale and $\tau_{\mathrm{age}}$ is the system age. As briefly introduced earlier, we delineate systems as 
\emph{dynamically young} if $\tau_{\mathrm{cir}} > \tau_{\mathrm{age}}$ 
and \emph{dynamically old} if $\tau_{\mathrm{cir}} < \tau_{\mathrm{age}}$ (as defined in Q14). \emph{Dynamically young} planets should still have eccentric orbits and \emph{dynamically old} ones should already have circular orbits. Then, $Q_\mathrm{p}$ should take the value that maximizes the difference between the eccentricity distributions of dynamically young and old systems. Any dynamically young planets we find on circular orbits are disk migration candidates, since its circular orbit cannot be explained by tidal circularization required for HEM. Conversely, dynamically old planets on circular orbits are both compatible with disk migration and HEM.

To the simplest approximation of equilibrium tides with a constant phase lag\footnote{With alternative formulation assuming a constant time lag, tidal quality factor will be inversely proportional to tidal forcing frequency, \( Q_\mathrm{p} \propto 1/|\omega_\mathrm{f}| \). Such formulation is more appropriate for fluid planets \citep{Hut_1981,Eggleton_1998,Mardling_2002}. We nonetheless opt for the constant phase lag model, which allows for a population level estimate of a constant \( Q_\mathrm{p} \).}, the circularization timescale of a planet can be expressed as a function of the tidal quality factor, \( Q_\mathrm{p} \), as follows:

\begin{equation}
\begin{split}
\tau_{\mathrm{cir}} &= 1.6\,\mathrm{Gyr} 
    \left(\frac{Q_{\mathrm{p}}}{10^6}\right) 
    \left(\frac{M_{\mathrm{p}}}{M_{\mathrm{J}}}\right) \\
    &\quad \times \left(\frac{M_*}{M_{\odot}}\right)^{-1.5} 
    \left(\frac{R_{\mathrm{p}}}{R_{\mathrm{J}}}\right)^{-5} 
    \left(\frac{a}{0.05\,\mathrm{AU}}\right)^{6.5}
\end{split}
\label{tau_circ}
\end{equation}

\noindent where \( M_{\mathrm{p}} \), \( M_{\mathrm{*}} \), \( R_{\mathrm{p}} \), and \( a \) represent the planet mass, stellar mass, planet radius, and orbital semi-major axis, respectively.\footnote{The equation assumes that (1) tidal dissipation in the planet is dominant in eccentricity damping and (2) the planets are tidally locked. The first assumption allows us to ignore the tidal dissipation in the star, for which we would need to incorporate another term including $Q_\mathrm{*}$, which is another uncertain variable. This assumption is valid as $|\dot e_\mathrm{p}/\dot e_\mathrm{*}| \sim R_*/R_\mathrm{p} \sim 10$ when the planet and the star has similar mean densities, as in the case of Sun and Jupiter. The second assumption allows us to reduce $Q_\mathrm{p}$ to a constant value, which would otherwise be a function of planet mean motion and spin frequency \citep{Tremaine_2023}. Most hot Jupiters are expected to be tidally locked.
}

It is important to note that dynamical tides become dominant in the circularization at higher eccentricities, which is particularly relevant in the context of HEM, and the above approximation does not capture this effect. It has been derived that, dissipation becomes more efficient with larger eccentricities \citep{Wisdom_2008}, and especially during the super-eccentric phase (\( e > 0.9 \)), damping due to dynamical tides should be extremely rapid \citep{Wu2018,Vick_Lai_Anderson_2019}. This is consistent with the observational scarcity of super-eccentric Jupiters \citep{Dawson_2015}. As a result, Equation \ref{tau_circ} likely overestimates $\tau_\mathrm{cir}$ for systems with non-zero eccentricity.

However, we can at least say that we are not overestimating $\tau_\mathrm{cir}$ for planets observed on circular or nearly circular orbits: if a planet attained its circular orbit through significant eccentricity damping, it must be true that equilibrium tides dominated in the final stages at lower eccentricities on top of dynamical tides at higher eccentricities. It must then be true that it would have at least taken $\tau_\mathrm{cir}$ calculated from Equation \ref{tau_circ} with the currently observed parameters on top of any prior circularization due to dynamical tides. This implies that we are safe from wrongfully labeling some planets as disk migration candidates with this equilibrium tides approximation. On the other hand, it is possible that $\tau_\mathrm{cir}$ is underestimated, in which case we could be overlooking planets that would otherwise satisfy \( \tau_\mathrm{cir} > \tau_\mathrm{age} \) when dynamical tides are accounted for. However, this is a relatively minor concern compared to the risk of wrong labeling. 

Using Equation \ref{tau_circ}, we optimize the value of \( Q_\mathrm{p} \) by considering its range from \( 10^4 \) to \( 10^7 \), divided into 40 bins. For each value of \( Q_\mathrm{p} \) within a bin, we repeat the following procedure 3,000 times:

\begin{enumerate}
    \item \textbf{Sample system parameters}: To calculate \( \tau_{\mathrm{cir}} \), we sample \( M_{\mathrm{p}} \), \( M_{\mathrm{*}} \), \( R_{\mathrm{p}} \), and \( a \) for each planet. These parameters are sampled from either (1) a normal distribution with the reported median as the mean and the larger of the uncertainties as the standard deviation, or when only an upper limit is available, (2) a uniform distribution with bounds set by the reported upper limit and the smallest median value observed across all planets in our sample. Since using the larger of the uncertainties in the normal distribution can sometimes result in negative sampled values, we also apply the same lower bound to the normal distribution to ensure that the values remain physically meaningful.

    \item \textbf{Sample system age \( \tau_{\mathrm{age}} \)}: To compute \( \tau_{\mathrm{cir}}/\tau_{\mathrm{age}} \), we sample \( \tau_{\mathrm{age}} \) using the same strategy outlined above. We then classify the planets into dynamically young and old groups.

    \item \textbf{Sample eccentricity}: We sample eccentricity using the same strategy. As previously mentioned in Section \ref{sec:sample_selection}, planets on assumed circular orbits are omitted.

    \item \textbf{Kolmogorov–Smirnov (KS) test}: We use the KS test to evaluate how strongly the eccentricity distributions of dynamically young and old groups differ, treating the $p$-value as a relative measure of separation against other trial $Q_\mathrm{p}$ values rather than an absolute significance test.
\end{enumerate}

The result of each iteration is the $Q_\mathrm{p}$ that yields the minimum $p$-value. We emphasize that the absolute $p$-values are not meaningful, but their relative variation across $Q_\mathrm{p}$ are: a single KS test for a single $Q_\mathrm{p}$ is not a measure of the difference between the eccentricity distribution of dynamically young and old planets, but the collection of $p$-values for varying $Q_\mathrm{p}$ serves as a relative indicator and provides a heuristic for locating the ``best-fit'' $Q_\mathrm{p}$ and thus the circularization boundary. To construct a posterior distribution for $Q_\mathrm{p}$, we resample these ``best-fit'' values in each iteration.

\section{Results} \label{sec:results}
\subsection{Inferred $Q_\mathrm{p}$}\label{sec:inferred_qp}

\begin{figure}
  \centering
  \includegraphics[width=\linewidth]{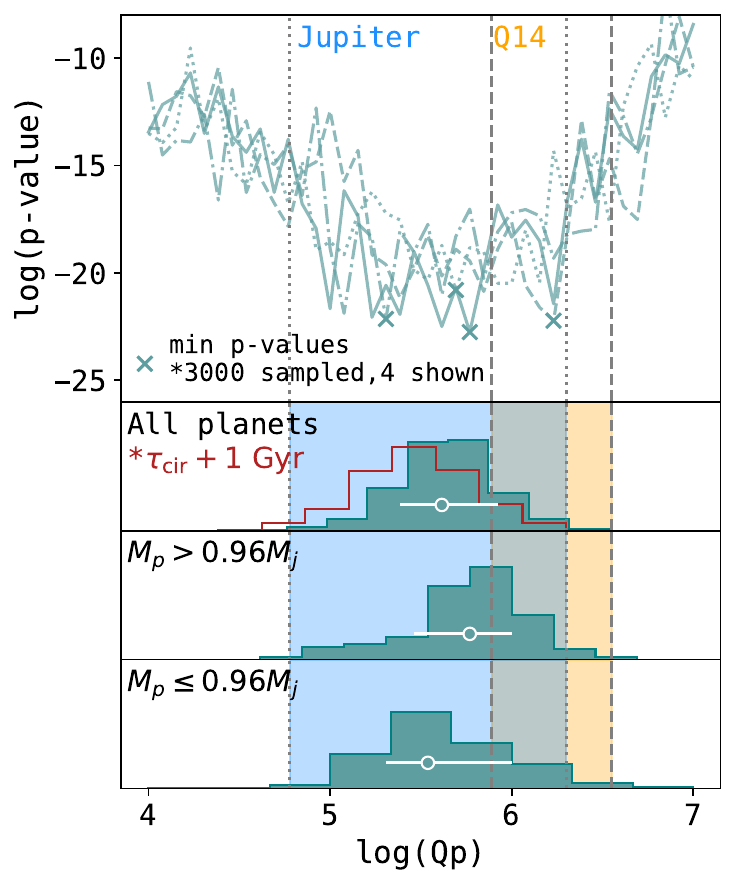}
  \caption{\textit{Top:} $p$-values from KS tests with trial \( Q_\mathrm{p} \) comparing the eccentricity distributions of dynamically old and young planets across four representative iterations (out of 3,000 total), distinguished by different line styles. In each curve, the minimum $p$-value among the trial \( Q_\mathrm{p} \) values in the same iteration is marked with an ``x''. \textit{Bottom three panels:} Histograms of $Q_\mathrm{p}$ corresponding to the locations of minimum $p$-values in each of the 3,000 iterations, shown for (1) all planets, (2) planets above the median mass ($M_\mathrm{p} > 0.96\,M_\mathrm{J}$), and (3) those below. In the first panel, we also overlay the case with a 1 Gyr offset applied to $\tau_\mathrm{cir}$, simulating the delay introduced by preceding processes such as eccentricity excitation. Dotted lines and blue shading indicate the constraints on $Q_\mathrm{J}$, while dashed lines and orange shading show constraints on $Q_\mathrm{p}$ from \citet{Quinn+2014}.
  }

\label{qp_fit_all_case}
\end{figure}

\begin{figure*}
  \centering
  \includegraphics[width=\linewidth]{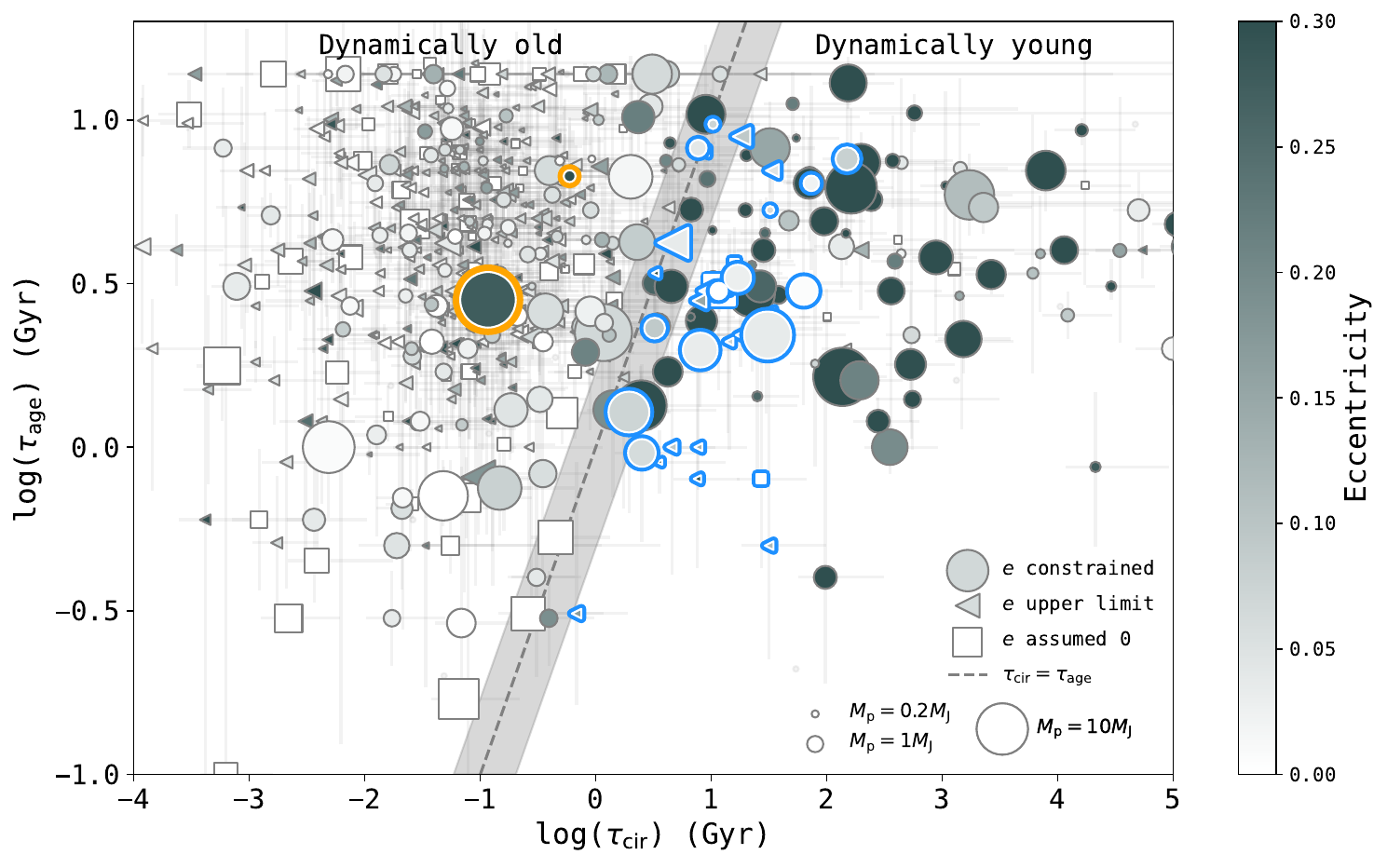}
  \caption{A recreation of Figure 4 in \citet{Quinn+2014}. The system age $\tau_\mathrm{age}$ is plotted against the tidal circularization timescale $\tau_\mathrm{cir}$ in logarithmic space, with color contours indicating orbital eccentricity. The dashed line marks where $\tau_\mathrm{cir} = \tau_\mathrm{age}$, based on $Q_\mathrm{p}=4.9\times10^5$. The shaded region reflects the uncertainty in $Q_\mathrm{p}$. Marker size corresponds to planetary mass, and planets with only mass upper limits are excluded. Planets with fully constrained eccentricities are shown as circles, those with upper limits are triangles, and assumed circular orbits are indicated by squares. Dynamically old but eccentric outliers are highlighted in orange, while disk migration candidate hot Jupiters ($a<0.1$ AU) are highlighted in blue. We note that $(\tau_\mathrm{cir}/\tau_\mathrm{age})_\mathrm{median} > 1$ does not strictly imply $\tau_{\mathrm{cir,\,median}}/\tau_{\mathrm{age,\,median}} > 1$, and vice versa. This results in one planet to the left of the dashed line being highlighted in blue, and another to the right not being highlighted.
}
\label{tau_age_plot}
\end{figure*}

Figure~\ref{qp_fit_all_case} summarizes the result of this experiment, revealing a minimum in the \( p \)-value (among other trial \( Q_\mathrm{p} \) in the same iteration) around \( Q_\mathrm{p} \sim 5 \times 10^5 \). Four of the 3,000 iterations are shown for visual clarity. With the procedure outlined in Section \ref{sec:constraining_qp}, we obtain a median and 68\% confidence interval of \( Q_\mathrm{p} = 4.9^{+3.5}_{-2.5} \times 10^5 \), which should be interpreted as a typical value for a hot Jupiter, and actual values may vary for individual planets. Nevertheless, this value is fully consistent with the range estimated for Jupiter in our own Solar System (\( 6 \times 10^4 < Q_\mathrm{p} < 2 \times 10^6 \); \citealt{YoderPeale_1981}), indicated by the blue shaded region in Figure~\ref{qp_fit_all_case}. This agreement may suggest that typical hot Jupiters share similar internal structures with Jupiter.

$Q_\mathrm{p}$ by Q14 of $8\times10^5<Q_\mathrm{p}<3\times10^6$ (shaded in orange) is slightly larger than both our result and Jupiter's value. We interpret the discrepancy with our refined value as a result of increased sample size, as well as sample selection limiting to transiting planets with measured mass, radius and eccentricity. In Q14, the radii of some planets were inferred using a mass-radius relation, which carries substantial uncertainty for Jovian-mass planets \citep{Weiss_2013}. Additionally, some systems were assumed to have zero eccentricity. Both assumptions can bias the estimation of \( Q_\mathrm{p} \), particularly when applied for planets near the circularization boundary at \( \tau_{\mathrm{cir}} \sim \tau_{\mathrm{age}} \).

A complication in our estimation of $Q_\mathrm{p}$ (and in Q14) is that eccentricity may be excited on secular timescales prior to circularization, particularly through the von Zeipel–Lidov–Kozai mechanism \citep{von_Zeipel_1910,Lidov_1962,Kozai_1962,Naoz_2016}, one of the most promising channels for HEM. Periods of dynamical stability may also precede such excitation events \citep{Weldon_2025}. Because our method adjusts $Q_\mathrm{p}$ without accounting for these earlier processes, the inferred $Q_\mathrm{p}$ could be systematically overestimated, effectively ``absorbing'' the delay while the true $\tau_\mathrm{cir}$ is shorter and dissipation more efficient. To test this possibility, we repeated the analysis with a constant 1~Gyr offset applied to $\tau_\mathrm{cir}$ (excluding systems younger than 1~Gyr). As we show in Figure \ref{qp_fit_all_case} this treatment shifts the median $Q_\mathrm{p}$ slightly in the expected direction (i.e., toward more efficient dissipation), but does not alter its overall order of magnitude.

Related to this, some systems with $\tau_\mathrm{cir} \ll \tau_\mathrm{age}$ are observed at high eccentricities, which may reflect recent arrival at their current orbits owing to unusually long excitation timescales (see also Section~\ref{Dynamically_old_but_eccentric}). Since we observe only two such cases, however, it seems unlikely that excitation timescales preceding circularization vary substantially among hot Jupiters; otherwise, the circularization boundary would be smeared out by the presence of many such systems, preventing this analysis altogether. We therefore conclude that while our $Q_\mathrm{p}$ estimate may be slightly overestimated, it remains robust at the order-of-magnitude level.

Figure \ref{tau_age_plot} is a recreation of Figure 4 in Q14, where we can see that that the optimization of $Q_\mathrm{p}$ is successful such that we see most eccentric planets occurring to the right of the dashed line at $\tau_{\mathrm{cir}}\sim\tau_{\mathrm{age}}$. We note that the interpretation of planets with small $\tau_{\mathrm{cir}}$ needs some caution. Planets with small $\tau_{\mathrm{cir}}$ naturally have small semi-major axis, and it is likely that their radius expanded post-migration \citep{Thorngren_2024}. As $\tau_{\mathrm{cir}}$ scales to the fifth power of planet radius,  $\tau_{\mathrm{cir}}$ calculated for these planets might not be accurate. However, these planets with the orbital periods of 1-2 days would always lie far from the $\tau_{\mathrm{cir}}\sim\tau_{\mathrm{age}}$ line, and have negligible impact in our estimation of \( Q_\mathrm{p} \).

As mentioned earlier, $Q_\mathrm{p}$ derived by this method does not represent the universal value across all Jovian mass planets, but should be treated as a typical or an average value of a Jovian mass planet. This has also been pointed out by Q14, and why some of the previous works on $Q_\mathrm{p}$ choose to constrain it for individual planets rather than as a population. In reality, $Q_\mathrm{p}$ likely depends on the internal structure, temperature, rotation and other unique properties of each planet. To account for this uncertainty as best possible within our framework, we repeat the same experiment to constrain \( Q_\mathrm{p} \), this time dividing the sample into planets with masses above and below the median value of \( 0.96\,M_\mathrm{J} \).

Interestingly, we find that the inferred values of \( Q_\mathrm{p} \) for the subsamples above and below \( 0.96\,M_\mathrm{J} \) are consistent within \(1\sigma\) (see bottom panels of Figure \ref{qp_fit_all_case}). However, the median value for the lower-mass subsample is marginally smaller, a shift consistent with the theoretical prediction that tidal dissipation is more efficient in smaller planets \citep{Goldreich_Soter_1966}. When we tried further dividing the samples by 25th, 50th, 75th quantiles, we find that there are not enough samples around the $\tau_{\mathrm{cir}}\sim\tau_{\mathrm{age}}$ boundary in the two intermediate mass bins to gain meaningful constraints on $Q_\mathrm{p}$. The lack of samples in these intermediate mass bins might be related to their efficiency of migration, which we explore in Section \ref{sec:discussion}.

Given that we include sub-Saturnian mass planets in our sample, we also compare the value of our $Q_\mathrm{p}$ to that of Saturn in our Solar System, $Q_\mathrm{s}$. Similarly to the case in Jupiter, using equilibrium tides to explain the position of satellites require $Q_\mathrm{s}\sim10^{4-6}$ \citep{Goldreich_Soter_1966,Sinclair_1983}, which is consistent with our results. However, the migration rate of Titan from its astrometry yields $Q_\mathrm{s}\sim100$, differing from theoretical expectations by orders of magnitude \citep{Lainey_2012}. It is proposed that a resonance lock between the oscillation mode of Saturn and its moons' tidal forcing frequency is responsible for such rapid dissipation and migration, in which case our derived $Q_\mathrm{p}$ and $Q_\mathrm{s}$ are not directly comparable \citep{Fuller_2016}. 

We conclude by also drawing comparisons to $Q_\mathrm{p}$ constrained for exoplanets of even smaller mass, namely the Neptune-mass GJ 436b \citep{Butler_2004} and super-Earth GJ 876d \citep{Marcy_1998}. $Q_\mathrm{p} = 2\times10^5-10^6$ is inferred for GJ~436b in order for tidal heating to explain the empirically derived interior temperature $T_\mathrm{int} \sim 300–350$K \citep{Morley_2017}. For GJ~876d, maintaining its forced eccentricity due to the Laplace-like resonance with its two companion planets, against ongoing tidal damping, requires $Q_\mathrm{p} \sim 10^{4}$–$10^{5}$. $Q_\mathrm{p}$ derived of exoplanets so far hence point, somewhat regardless of their mass, to values similar to that of Solar System gas giants.

\subsection{Disk migration candidates}\label{sec:candidates}
\begin{figure*}
  \centering
  \includegraphics[width=\linewidth]{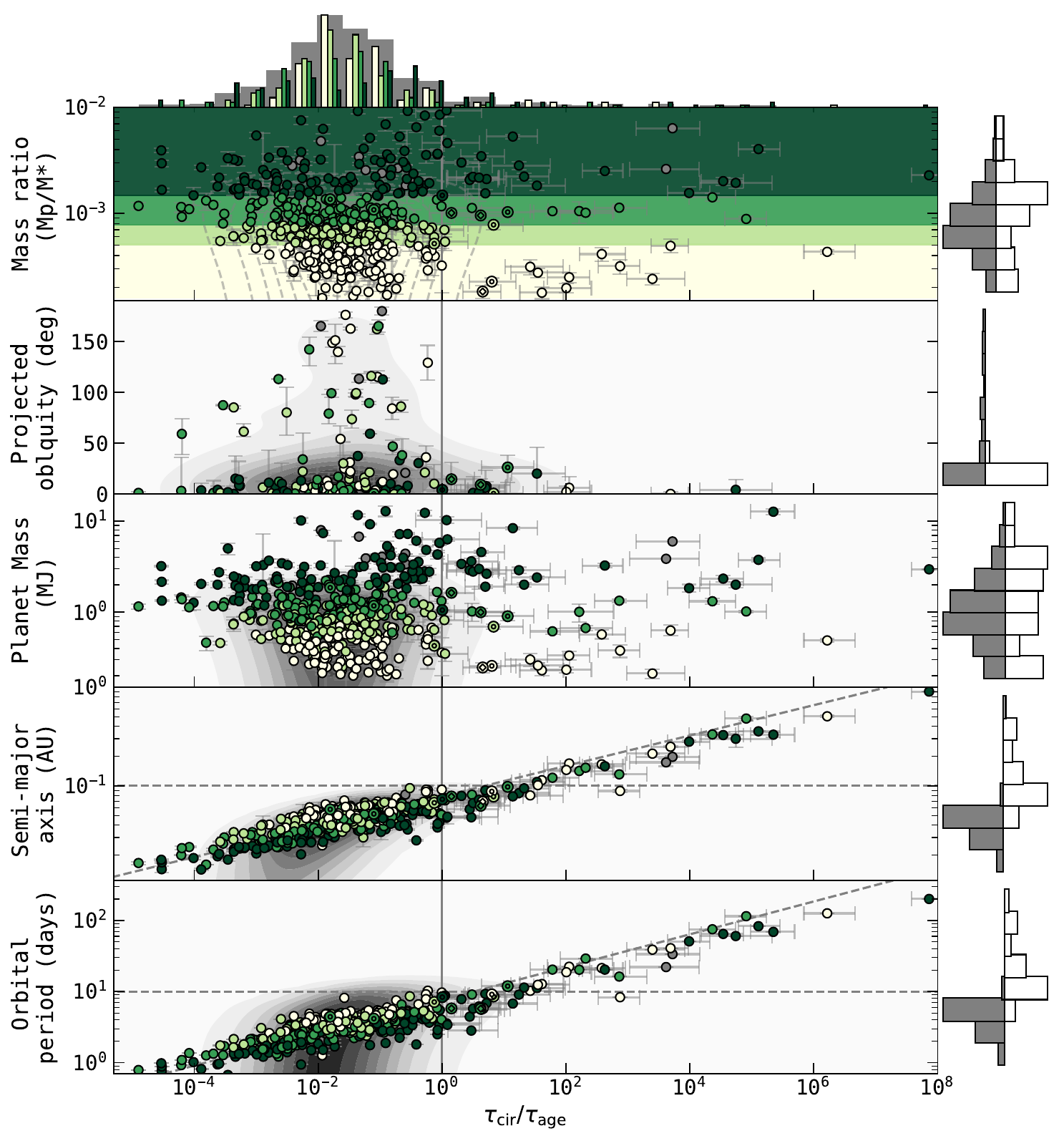}
  \caption{System parameters as a function of dynamical age, \( \tau_{\mathrm{cir}} / \tau_{\mathrm{age}} \), for planets with \( e < 0.1 \) (including those on assumed circular orbits). The vertical line marks the threshold for disk migration candidates. Points are color-coded by mass ratio in quartiles, with four shades of green: darkest for the top 25\% (75–100th percentile), medium-dark for 50–75th, medium-light for 25–50th, and lightest for 0–25th. Gray points denote systems with only upper limits on mass ratio. Double-filled circles indicate nearby companions, and circles filled with diamonds indicate systems with known binary companions. Gray and white histograms show distributions for dynamically old and young systems, respectively. For clarity, error bars on dynamical age are shown only for disk migration candidates. In the semi-major axis and orbital period panels, horizontal dashed lines mark \( a = 0.1 \) AU and \( P = 10 \) days; diagonal lines trace expected dynamical age assuming median system values. The entire plot or individual panels can be regenerated with \texttt{ident-dm} (\url{https://github.com/ykawai65/ident-dm}) also archived in Zenodo \citep{ident_dm}.}
  \label{tau_age_plot_alt}
\end{figure*}

We plot $\tau_\mathrm{cir}/\tau_\mathrm{age}$ of all planets in our sample on circular orbits ($e<0.1$) in Figure \ref{tau_age_plot_alt}. Planets to the right of the vertical line at $\tau_\mathrm{cir}/\tau_\mathrm{age}=1$ are dynamically young planets, and hence disk migration candidates unexplained by HEM (i.e. $\tau_\mathrm{cir}>\tau_\mathrm{age} ~\&~ e<0.1$). We plot these systems over relevant parameters including mass ratio, projected obliquity, planet mass, and orbital separation. If we adopt the conventional but inclusive definition of hot Jupiters ($0.2\,M_\mathrm{J} < M_\mathrm{p} < 13\,M_\mathrm{J}$ and $a < 0.1$\,AU), our sample contains 13 disk migration candidate hot Jupiters with fully constrained eccentricities. When we include additional candidates with either upper limits on eccentricity or assumed circular orbits, the sample expands to 36. These systems are listed in Table~\ref{ident-dm}. The full table of dynamical age and other relevant parameters for all planets in the sample is also available on Github\footnote{\url{https://github.com/ykawai65/ident-dm}} and slo archived in Zenodo \citep{ident_dm}.

\subsection{Dynamically old but eccentric systems}
\label{Dynamically_old_but_eccentric}
In Figure~\ref{tau_age_plot}, we also identify two notable outliers on the left side of the diagram: XO-3~b (\( \tau_{\mathrm{cir}} / \tau_{\mathrm{age}} \sim 0.04,\ e\sim0.27 \); \citealt{Johns-Krull_2008}) and CoRoT-16~b (\( \tau_{\mathrm{cir}} / \tau_{\mathrm{age}} \sim 0.1,\ e\sim0.37 \); \citealt{Ollivier_2012}), highlighted in orange. Despite being dynamically old based on their short circularization timescales relative to their system ages, both planets exhibit significant orbital eccentricities, contrary to expectations. 

The discrepancy has also been discussed in the respective discovery papers by \citet{Johns-Krull_2008} and \citet{Ollivier_2012}. Potential explanations mentioned include: (1) inaccurate stellar age estimates; (2) unusual internal structures that result in inefficient tidal dissipation, particularly for XO-3~b, which is an exceptionally massive hot Jupiter with \( M_\mathrm{p} \sim 11.7\ M_\mathrm{J} \); and (3) recent excitation of eccentricity with ongoing circularization. Given the rarity of these outliers (less than 1\% of the population), it is plausible that these systems are indeed undergoing circularization and are observed in a transient phase. Another possibility is forced eccentricity due to undetected companions, although no such bodies have been identified in either system.

\begin{table*}[ht]
\centering
\small
\begin{tabular}{p{2.1cm} | l | l | l | l | l | l | l | l | l}
\hline
\multicolumn{10}{l}{\textbf{Candidates with eccentricity constrained}} \\
\hline
Planet & $M_\mathrm{p}$ [$M_\mathrm{J}$] & $M_\ast$ [$M_\odot$] & $\log q$ & $R_\mathrm{p}$ [$R_\mathrm{J}$] & $P$ [days] & $a$ [au] & $e$ & $\tau_\mathrm{age}$ [Gyr] & $\tau_\mathrm{cir}/\tau_\mathrm{age}$ \\
\hline
CoRoT-30* b & $2.9 \pm 0.22$ & $0.98_{-0.05}^{+0.03}$ & $-2.55$ & $1.01$ & $9.06005$ & $0.0844$ & $0.007 \pm 0.031$ & $3_{-2.4}^{+3.7}$ & $17.4_{-10.9}^{+38.2}$ \\
K2-114 b & $2.01 \pm 0.12$ & $0.863_{-0.031}^{+0.037}$ & $-2.65$ & $0.94$ & $11.39093$ & $0.0944$ & $0.079 \pm 0.03$ & $7.6_{-4.6}^{+4.2}$ & $21.1_{-11.1}^{+32.0}$ \\
KELT-24 b* & $4.59 \pm 0.18$ & $1.268 \pm 0.063$ & $-2.46$ & $1.13$ & $5.55149$ & $0.0665$ & $0.0319_{-0.0074}^{+0.0079}$ & $1.98_{-0.79}^{+0.95}$ & $4.3_{-2.3}^{+5.8}$ \\
Kepler-14 b & $8.4_{-0.34}^{+0.35}$ & $1.512 \pm 0.043$ & $-2.28$ & $1.14$ & $6.79012$ & $0.07826$ & $0.035 \pm 0.02$ & $2.2_{-0.1}^{+0.2}$ & $13.9_{-8.6}^{+21.0}$ \\
NGTS-9 b* & $2.9 \pm 0.17$ & $1.34 \pm 0.05$ & $-2.68$ & $1.07$ & $4.43527$ & $0.058$ & $0.06_{-0.052}^{+0.076}$ & $0.96 \pm 0.6$ & $2.8_{-1.6}^{+5.3}$ \\
TOI-628 b* & $6.33_{-0.31}^{+0.29}$ & $1.311_{-0.075}^{+0.066}$ & $-2.34$ & $1.06$ & $3.40957$ & $0.0486$ & $0.072_{-0.023}^{+0.021}$ & $1.28_{-0.91}^{+1.6}$ & $1.2_{-0.7}^{+2.8}$ \\
TOI-1130 c* & $1.057_{-0.016}^{+0.006}$ & $0.684_{-0.017}^{+0.016}$ & $-2.83$ & $1.16$ & $8.35019$ & $0.0731$ & $0.0398_{-0.0002}^{+0.0009}$ & $8.2_{-4.9}^{+3.8}$ & $1.0_{-0.5}^{+1.6}$ \\
TOI-2000 c* & $0.257_{-0.014}^{+0.015}$ & $1.082_{-0.05}^{+0.059}$ & $-3.64$ & $0.73$ & $9.12706$ & $0.0878$ & $0.063_{-0.022}^{+0.023}$ & $5.3 \pm 2.7$ & $6.3_{-3.6}^{+9.3}$ \\
TOI-2202 b* & $0.904_{-0.1}^{+0.087}$ & $0.84 \pm 0.03$ & $-2.99$ & $0.98$ & $11.91261$ & $0.09766$ & $0.022_{-0.015}^{+0.022}$ & $6.4_{-3.9}^{+4.4}$ & $11.4_{-7.1}^{+18.9}$ \\
TOI-3693 b* & $1.02_{-0.22}^{+0.24}$ & $0.867_{-0.037}^{+0.036}$ & $-2.95$ & $1.12$ & $9.08852$ & $0.0813$ & $0_{-0.052}^{+0.048}$ & $3_{-2.1}^{+3.8}$ & $3.0_{-2.0}^{+7.1}$ \\
TOI-5573 b & $0.35 \pm 0.06$ & $0.619 \pm 0.023$ & $-3.27$ & $0.87$ & $8.79759$ & $0.0712$ & $0.072_{-0.048}^{+0.067}$ & $9.7_{-2.7}^{+3.9}$ & $1.1_{-0.6}^{+1.5}$ \\
TOI-6029 b & $1.635 \pm 0.032$ & $1.548_{-0.085}^{+0.027}$ & $-3.00$ & $1.28$ & $5.7987$ & $0.07799$ & $0.091_{-0.015}^{+0.017}$ & $2.31_{-0.19}^{+0.58}$ & $1.4_{-0.8}^{+1.9}$ \\
WASP-38 b & $2.648_{-0.058}^{+0.057}$ & $1.2 \pm 0.04$ & $-2.68$ & $1.23$ & $6.87188$ & $0.08564$ & $0.0278_{-0.0028}^{+0.003}$ & $3.29_{-0.53}^{+0.42}$ & $5.2_{-2.9}^{+5.6}$ \\
\hline
\multicolumn{10}{l}{\textbf{Candidates with eccentricity upper limits}} \\
\hline
Planet & $M_\mathrm{p}$ [$M_\mathrm{J}$] & $M_\ast$ [$M_\odot$] & $\log q$ & $R_\mathrm{p}$ [$R_\mathrm{J}$] & $P$ [days] & $a$ [au] & $e$ & $\tau_\mathrm{age}$ [Gyr] & $\tau_\mathrm{cir}/\tau_\mathrm{age}$ \\
\hline
CoRoT-4 b & $0.703_{-0.073}^{+0.071}$ & $1.16_{-0.02}^{+0.03}$ & $-3.24$ & $1.19$ & $9.20205$ & $0.09$ & $< 0.14$ & $1_{-0.3}^{+1}$ & $6.9_{-4.5}^{+14.7}$ \\
CoRoT-6 b & $2.95 \pm 0.28$ & $1.05 \pm 0.05$ & $-2.57$ & $1.17$ & $8.88659$ & $0.0855$ & $< 0.18$ & $2.5_{-1.7}^{+2.1}$ & $11.4_{-6.8}^{+22.2}$ \\
CoRoT-8 b & $0.218_{-0.041}^{+0.033}$ & $0.88 \pm 0.04$ & $-3.63$ & $0.57$ & $6.21229$ & $0.063$ & $< 0.19$ & $3_{-3}^{+0}$ & $4.2_{-2.6}^{+9.3}$ \\
CoRoT-27 b & $10.3_{-0.83}^{+0.81}$ & $1.05 \pm 0.11$ & $-2.03$ & $1.01$ & $3.57532$ & $0.0476$ & $< 0.034$ & $4.21 \pm 2.72$ & $1.2_{-0.8}^{+3.2}$ \\
HATS-58 A b* & $1.03 \pm 0.23$ & $1.461 \pm 0.043$ & $-3.17$ & $1.09$ & $4.21809$ & $0.05798$ & $< 0.168$ & $0.31_{-0.2}^{+0.33}$ & $1.9_{-1.3}^{+4.4}$ \\
HATS-61 b* & $3.4 \pm 0.14$ & $1.076 \pm 0.014$ & $-2.52$ & $1.19$ & $7.81795$ & $0.07908$ & $< 0.092$ & $8.9_{-0.41}^{+0.31}$ & $2.1_{-1.0}^{+1.9}$ \\
Kepler-40 b & $2.07 \pm 0.33$ & $1.48 \pm 0.06$ & $-2.87$ & $1.17$ & $6.87349$ & $0.08$ & $< 0.15$ & $2.8 \pm 0.3$ & $2.8_{-1.4}^{+2.8}$ \\
Kepler-63 b* & $< 0.378$ & $0.98 \pm 0.04$ & $-3.43$ & $0.55$ & $9.43415$ & $0.08$ & $< 0.45$ & $0.21 \pm 0.05$ & $252.6_{-190.8}^{+404.8}$ \\
Kepler-74 b & $0.604_{-0.094}^{+0.088}$ & $1.18 \pm 0.04$ & $-3.31$ & $0.96$ & $7.34071$ & $0.0781$ & $< 0.32$ & $0.8_{-0.5}^{+0.9}$ & $8.2_{-4.7}^{+16.6}$ \\
Kepler-422 b & $0.49_{-0.12}^{+0.23}$ & $1.15 \pm 0.06$ & $-3.39$ & $1.15$ & $7.89145$ & $0.082$ & $< 0.55$ & $3.4_{-1.1}^{+0.9}$ & $1.0_{-0.7}^{+2.8}$ \\
TOI-622 b* & $0.303_{-0.072}^{+0.069}$ & $1.313 \pm 0.079$ & $-3.66$ & $0.82$ & $6.40251$ & $0.0708$ & $< 0.42$ & $0.9 \pm 0.2$ & $4.1_{-2.5}^{+5.8}$ \\
TOI-892 b* & $0.95 \pm 0.07$ & $1.28 \pm 0.03$ & $-3.15$ & $1.07$ & $10.62656$ & $0.092$ & $< 0.125$ & $2.2 \pm 0.5$ & $8.2_{-4.6}^{+9.0}$ \\
TOI-2158 b* & $0.82 \pm 0.08$ & $1.12 \pm 0.12$ & $-3.16$ & $0.96$ & $8.60077$ & $0.075$ & $< 0.07$ & $8 \pm 1$ & $1.1_{-0.6}^{+1.1}$ \\
WASP-59 b & $0.857_{-0.047}^{+0.046}$ & $0.72 \pm 0.04$ & $-2.94$ & $0.77$ & $7.91958$ & $0.0697$ & $< 0.12$ & $0.5_{-0.4}^{+0.7}$ & $48.9_{-35.1}^{+125.7}$ \\
WASP-84 b & $0.694_{-0.047}^{+0.049}$ & $0.85 \pm 0.06$ & $-3.11$ & $0.96$ & $8.52350$ & $0.0778$ & $< 0.077$ & $2.1 \pm 1.6$ & $6.8_{-4.1}^{+13.7}$ \\
WASP-99 b* & $2.8 \pm 0.13$ & $1.21 \pm 0.28$ & $-2.66$ & $1.02$ & $5.75251$ & $0.06692$ & $< 0.01$ & $3_{-0.5}^{+0.6}$ & $3.2_{-2.0}^{+5.0}$ \\
WASP-106 b* & $1.909 \pm 0.079$ & $1.18_{-0.07}^{+0.08}$ & $-2.81$ & $1.08$ & $9.28969$ & $0.0901$ & $< 0.046$ & $7 \pm 2$ & $5.0_{-2.7}^{+5.5}$ \\
WASP-129 b* & $1 \pm 0.1$ & $1 \pm 0.03$ & $-3.02$ & $0.93$ & $5.74815$ & $0.0628$ & $< 0.096$ & $1 \pm 0.9$ & $4.2_{-2.6}^{+8.3}$ \\
\hline
\multicolumn{10}{l}{\textbf{Candidates with assumed circular orbits}} \\
\hline
Planet & $M_\mathrm{p}$ [$M_\mathrm{J}$] & $M_\ast$ [$M_\odot$] & $\log q$ & $R_\mathrm{p}$ [$R_\mathrm{J}$] & $P$ [days] & $a$ [au] & $e$ & $\tau_\mathrm{age}$ [Gyr] & $\tau_\mathrm{cir}/\tau_\mathrm{age}$ \\
\hline
NGTS-33 b & $3.63 \pm 0.27$ & $1.6 \pm 0.11$ & $-2.66$ & $1.64$ & $2.82797$ & $0.048$ & $0$ & $0.03 \pm 0.02$ & $3.0_{-1.7}^{+5.1}$ \\
TOI-837 b & $0.379_{-0.061}^{+0.058}$ & $1.142_{-0.011}^{+0.008}$ & $-3.50$ & $0.82$ & $8.32491$ & $0.0888$ & $0$ & $0.04_{-0.00}^{+0.01}$ & $745.7_{-412.2}^{+811.9}$ \\
TOI-1775 b & $0.302 \pm 0.047$ & $0.922_{-0.015}^{+0.014}$ & $-3.50$ & $0.72$ & $10.24055$ & $0.08$ & $0$ & $0.8_{-0.5}^{+1.1}$ & $26.4_{-16.1}^{+62.5}$ \\
TOI-6038 A b & $0.247_{-0.031}^{+0.030}$ & $1.291_{-0.06}^{+0.066}$ & $-3.74$ & $0.57$ & $5.82673$ & $0.069$ & $0$ & $3.65_{-0.85}^{+0.92}$ & $4.5_{-2.4}^{+4.4}$ \\
EPIC 246851721~b & $3_{-1.2}^{+1.1}$ & $1.32 \pm 0.04$ & $-2.66$ & $1.05$ & $6.18023$ & $0.07229$ & $0$ & $3.02_{-0.46}^{+0.44}$ & $3.9_{-2.2}^{+4.5}$ \\
\hline
\end{tabular}
\caption{List of disk migration candidate hot Jupiters. Subsections correspond to targets with constrained eccentricity, upper limits, and those assumed to be on circular orbits. Asterisks denote Ariel Tier 3 targets.}
\label{ident-dm}
\end{table*}

\section{Discussion} \label{sec:discussion}
\subsection{Planets preferentially form primordially aligned}
\label{sec:primordial_alignment}
In this study, we find that obliquity in giant planets exhibits a clear cutoff at $\tau_\mathrm{cir}/\tau_\mathrm{age} \sim 1$ , a physically motivated threshold arising naturally from the requirement of HEM (see second panel in Figure \ref{tau_age_plot_alt}). Systems with $\tau_\mathrm{cir}/\tau_\mathrm{age} < 1$ —those dynamically old enough for tidal circularization to complete—show a wide range of obliquities, while systems with $\tau_\mathrm{cir}/\tau_\mathrm{age} > 1$ and on circular orbits are predominantly aligned. This trend holds across a wide range of planet masses, from Saturnian to Jovian, suggesting that planets form in primordially aligned protoplanetary disks. This implies that HEM is the primary driver of the observed high obliquities in planetary systems, with primordial misalignment playing only a minor confounding role when high obliquity is used to infer HEM.

This interpretation supports the earlier work by \cite{Rice_2022}, who proposed a cutoff in obliquity at $a/R_\ast \sim 11$ to distinguish between hot Jupiters ($a/R_\ast < 11$, by their definition), thought to migrate through violent channels, and warm Jupiters ($a/R_\ast > 11$), thought to migrate quiescently. The cutoff in $a/R_\ast$ is a strong argument for the primordial alignment of protoplanetary disks, since the tidal realignment timescale of obliquity (as opposed to orbital circularization) scales strongly with $a/R_\ast$. With little to no tidal realignment expected at large $a/R_\ast$, the observed alignment suggests that obliquity must have been low from the outset. Our results are fully consistent with their findings on the primordial alignment of protoplanetary disks. 

However, it should also be noted that employing $a/R_\ast$ as a boundary to distinguish migration pathways does not fully account for the presence of misaligned systems at larger separations—especially those with circular orbits. In fact, \cite{Rice_2022} noted a mixture of aligned and misaligned Saturn-mass planets beyond $a/R_\ast \sim 11$, which they attributed to increased susceptibility to scattering with decreasing planet mass. This explanation, however, implicitly requires distinct formation scenarios for Jupiter and Saturn mass planets on wider orbits. By contrast, the $\tau_{\mathrm{cir}}/\tau_{\mathrm{age}}$ framework offers a much simpler explanation. The tidal circularization timescale, rather than $a/R_\ast$ alone, separates dynamically violent from quiescent origins of giant planets on short orbits. For example, HAT-P-12~b has $a/R_\ast \sim 12$ and a circular orbit with significant obliquity. Yet it has $\tau_{\mathrm{cir}}/\tau_{\mathrm{age}} \sim 0.02$, fully consistent with a violent origin for this hot Saturn. 

We also note that rare cases of primordial misalignment exist, such as Kepler-56 and K2-290, which host coplanar planets that are both misaligned with respect to the stellar spin \citep{Huber_2013,Hjorth_2021}. In Figure~\ref{tau_age_plot_alt}, we find two systems consistent with non-zero obliquity that host binary companions: 
WASP-129\,b ($\lambda = 9^{\circ} \pm 6^{\circ}$; \citealt{Zak_2025}) and 
TOI-6029\,b ($\lambda = -14.4^{\circ\,+16.7^{\circ}}_{\,\,\,  -12.7^{\circ}}$; \citealt{Saunders_2024}). 
Although the sample size is small, both cases are consistent with a scenario in which binary companions misalign the protoplanetary disk \citep[e.g.][]{Batygin_2013}. 
The other misaligned planet, TOI-2202\,b ($\lambda = 26^{\circ\,+12^{\circ}}_{\,\,\,  -15^{\circ}}$; \citealt{Rice_2023}), harbors a near-resonant planetary companion, suggesting that low-level dynamical excitation rather than primordial misalignment may explain its observed obliquity \citep{Rice_2023}.

Additionally, the apparent decrease of misaligned planets with smaller $\tau_{\mathrm{cir}}/\tau_{\mathrm{age}}$ is qualitatively consistent with the hypothesis of orbital realignment, because planets with the shortest $\tau_{\mathrm{cir}}$ have the longest remaining to realign its orbit, and the realignment timescale is orders of magnitude larger than $\tau_{\mathrm{cir}}$. On a similar note, \cite{Rice_2022b} showed that planets with the shortest realignment timescale tend to be significantly more aligned. However, we note that interpretation of small $\tau_{\mathrm{cir}}$ in our work requires some caution for the reason outlined in Section \ref{sec:inferred_qp}.

Overall, because we began our experiment agnostic about the obliquity of planets in the sample, this cutoff also supports the validity of dynamical age as a proxy for disk migration. The presence of eccentric and misaligned systems beyond $\tau_{\mathrm{cir}}/\tau_{\mathrm{age}} \sim 1$ also proves that we are successfully isolating hot Jupiters that arrived via disk migration. 

\subsection{Nearby companions are preferentially found around dynamically young hot Jupiters}
\label{discussion_nearby_comp}

\begin{figure}
  \centering
  \includegraphics[width=\linewidth]{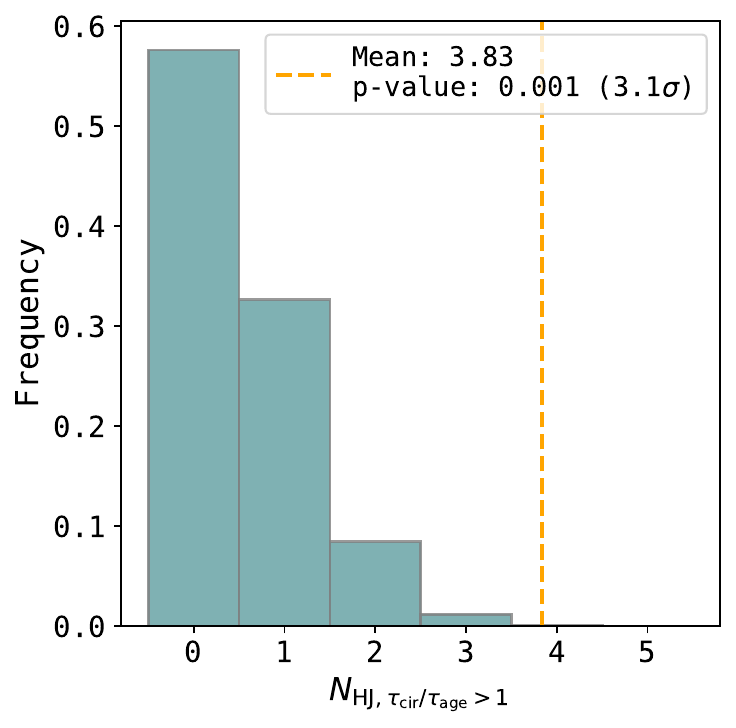}
  \caption{The result of the Monte Carlo experiment described in Section~\ref{discussion_nearby_comp}. While the sample of seven hot Jupiters with nearby companions contains an average of 3.83 dynamically young systems, a sample of seven randomly drawn hot Jupiters on circular orbits typically contains none. The resulting $p$-value is 0.001, corresponding to \( 3.1\sigma \).}
  \label{n_comp_test}
\end{figure}

Secondly, we observe a trend suggesting that the rare nearby companions ($P<50$ days and $R_\mathrm{p}>2R_\oplus$, adopted from \citealt{Huang_2016}) are more preferentially found around dynamically young hot Jupiters on circular orbits, which likely consist exclusively of ones that arrived via disk migration. The full sample of hot Jupiters with nearby companions in our study consists of WASP-47~b, WASP-84~b, WASP-132~b, TOI-1130~b, TOI-1408~b, TOI-2000~b and TOI-2202~b\footnote{All systems except TOI-2202~b host an inner companion, and TOI-2202~b hosts and outer companion ($P\sim24$ days). WASP-47~b hosts both inner and an outer companion. Kepler-730~b lacks a mass measurement, and is therefore excluded.} 

We observe in Figure \ref{tau_age_plot_alt} that while hot Jupiters in general exhibit a clustering around $\tau_{\mathrm{cir}}/\tau_{\mathrm{age}} \approx 0.01$, five out of the seven hot Jupiters with nearby companions lie beyond $\tau_{\mathrm{cir}}/\tau_{\mathrm{age}} \sim 0.1$, and three of them robustly exceed $\tau_{\mathrm{cir}}/\tau_{\mathrm{age}} > 1$. This concentration of nearby companions among dynamically young hot Jupiters on circular orbits points to a formation pathway that is compatible with companion survival, most plausibly, disk migration. This is yet another support for dynamical age as proxy for disk migration.

To assess whether this spread in $\tau_{\mathrm{cir}}/\tau_{\mathrm{age}}$ for hot Jupiters with nearby companions could arise by chance, we perform a Monte Carlo experiment. First, we draw a $\tau_{\mathrm{cir}}/\tau_{\mathrm{age}}$ value for each of the seven systems with nearby companions based on their posterior distributions. Then, we randomly select seven planets from the sample of hot Jupiters on circular orbits ($0.2 M_\mathrm{J}<M_\mathrm{p}<13 M_\mathrm{J},\ a<0.1$ AU, and $e<0.1$) and draw $\tau_{\mathrm{cir}}/\tau_{\mathrm{age}}$ values from their respective distributions. Repeating this process 100{,}000 times, we count how often the randomly drawn sets contain as many or more planets with $\tau_{\mathrm{cir}}/\tau_{\mathrm{age}} > 1$ as in the actual nearby companion sample. We find a $p$-value of 0.001 ($\sim3.1\sigma$) for this occurrence, indicating that such a spread in $\tau_{\mathrm{cir}}/\tau_{\mathrm{age}}$ is unlikely to happen by chance. We also show the result of this experiment in Figure \ref{n_comp_test}.

However, we should also consider if this result is due to any bias. The first possibility is the observational bias for the detection of TTV signals. As mentioned by \cite{Wu_2023}, the SNR of TTV signals scale with the orbital period as $\mathrm{SNR}_\mathrm{TTV} \sim P^{5/6}$, which indicates that there is a detection bias for TTVs of planets on longer orbits. Interestingly, two of the three current discoveries of nearby companions from TTVs come from systems with the shortest orbital periods (TOI-1408~b: $P\sim$ 4.4 days and WASP-47~b: $P\sim$ 4.2 days). Other discoveries are originally from transits of the nearby companions. The current samples therefore shows a trend opposite to what would be expected from observational bias.

Additionally, we consider the fact that dynamically old hot Jupiters tend to have shorter semi-major axis compared to dynamically young ones, causing mutual Hill radii between them and the companions to also be smaller. In other words, the orbits of dynamically old hot Jupiters are naturally more likely to leave the companion (inner in this case) with insufficient stable orbital space. Hence, the observed trend might simply reflect such physical constraint rather than migration history.

The Hill stability criterion for low eccentricities ($e\leq 0.1$) is written as

\begin{equation}
    \frac{a_\mathrm{out}}{a_\mathrm{in}} >2.4(\mu_\mathrm{in}+\mu_\mathrm{out})^{1/3}+1,
\end{equation}

\noindent where $a$ and $\mu$ with the respective subscripts denote the semi-major axis and the planet-to-star mass ratio of the inner and outer planets respectively \citep{Gladman_1993}. For a Jupiter sized planet and an Earth like planet around a sun like star, $a_\mathrm{J} > 1.24a_\mathrm{comp}$ is required for the planets to be Hill stable. All of our hot Jupiters with inner companions satisfy this criteria, but some dynamically old hot Jupiters will require a hypothetical inner companion to reside on extremely short orbits to be Hill stable, at which point the planet is likely tidally disrupted by the host star. We therefore repeat the same experiment, while excluding hot Jupiters with $a_\mathrm{J} <1.24\times0.024$ au, where 0.024 is the semi-major axis of hypothetical inner companion deduced from the smallest of the currently discovered companions (WASP-84c: \citealt{Maciejewski_2023}). We find a $p$-value of 0.002 ($\sim2.9\sigma$), which suggests that dynamically young hot Jupiters are more likely to host inner (and outer) companions even when the samples are limited to hot Jupiters that can host inner companions in the first place.

On the other hand, since HEM is incompatible with the survival of nearby companions, the presence of at least three to four dynamically old hot Jupiters with nearby companions (TOI-1130~b being the edge case with $\tau_\mathrm{cir}/\tau_\mathrm{age}\sim1$, see Table \ref{ident-dm}) must be explained through disk migration. This observation constrains the combination of two key parameters: the disk migration fraction among hot Jupiters, \( f_\mathrm{disk} \), and the disruption rate of nearby companions during disk migration, \( f_\mathrm{disrupt} \). If disruption were negligible (low \( f_\mathrm{disrupt} \)), finding only four such systems would be extremely unlucky unless disk migration is intrinsically rare (low \( f_\mathrm{disk} \)). Conversely, if disruption were very efficient (high \( f_\mathrm{disrupt} \)), finding even four surviving systems would be extremely lucky even if almost all hot Jupiters formed via disk migration (high \( f_\mathrm{disk} \)). Thus, even though the two parameters are degenerate, the observed number of dynamically old hot Jupiters with nearby companions suggests either that HEM is the dominant formation channel of hot Jupiters, or that disk migration also disrupts a moderate fraction of nearby companions.

Assuming all nearby companions originate from an initial fraction \( f_\mathrm{comp} \), and that only a fraction \( 1 - f_\mathrm{disrupt} \) survive disk migration, the disk migration fraction can be expressed as:

\begin{align}
    f_\mathrm{disk} &= \frac{N_\mathrm{comp,old}}{(1 - f_\mathrm{disrupt}) \times f_\mathrm{comp} \times N_\mathrm{old}},
\end{align}

\noindent where \( N_\mathrm{old} \) is the number of dynamically old hot Jupiters on circular orbits in the sample, and \( N_\mathrm{comp,old} \) is the subset of those with nearby companions.

Adopting \( f_\mathrm{disrupt} = 0.99 \) based on the simulations by \cite{He_Wu_2024}, along with a conservative assumption of \( f_\mathrm{comp} = 1 \) (i.e., all systems initially host companions), and using \( N_\mathrm{old} = 465 \) from our sample, we obtain

\begin{align}
    f_\mathrm{disk} = \frac{4}{0.01 \times 1 \times 465} \approx 0.86.
\end{align}

That is, 86\% of dynamically old hot Jupiters would have to originate from disk migration to account for the four surviving nearby companions under a 99\% disruption rate. This is an implausibly high fraction given the observed prevalence of high stellar obliquities typically interpreted as signatures of HEM.

In reality, the true number of nearby companions is likely higher once detection incompleteness is taken into account, and the initial companion fraction is probably less than unity. Thus, the required \( f_\mathrm{disk} \) would be even larger, potentially exceeding 100\%, which is unphysical. This makes a very high disruption rate (\( f_\mathrm{disrupt} \sim 99\% \)) unlikely in the context of disk migration. Nevertheless, more moderate disruption rates, such as the 2/3 proposed by \cite{Wu_He_2023}, remain consistent with the current observations when paired with a more modest \( f_\mathrm{disk} \).

\subsection{A dip in mass ratio distribution of disk migration candidates}
\label{sec:mass_ratio}

In Figure \ref{tau_age_plot_alt}, we observe a possible dip in the mass ratio histogram of dynamically young planets on circular orbits (i.e. disk migration candidates) around \( 3 \times 10^{-4} < q < 8 \times 10^{-4} \), as opposed to the ubiquity of dynamically old counterparts in this mass ratio range. If this dip is real, it is unlikely due to observational biases since systems with smaller mass ratios are detected. Interestingly, the location of the dip in occurrence is the same as what was recently suggested for planets on wider orbits with microlensing survey \citep{Zang_2025}. 

\begin{figure*}
  \centering
  \includegraphics[width=\linewidth]{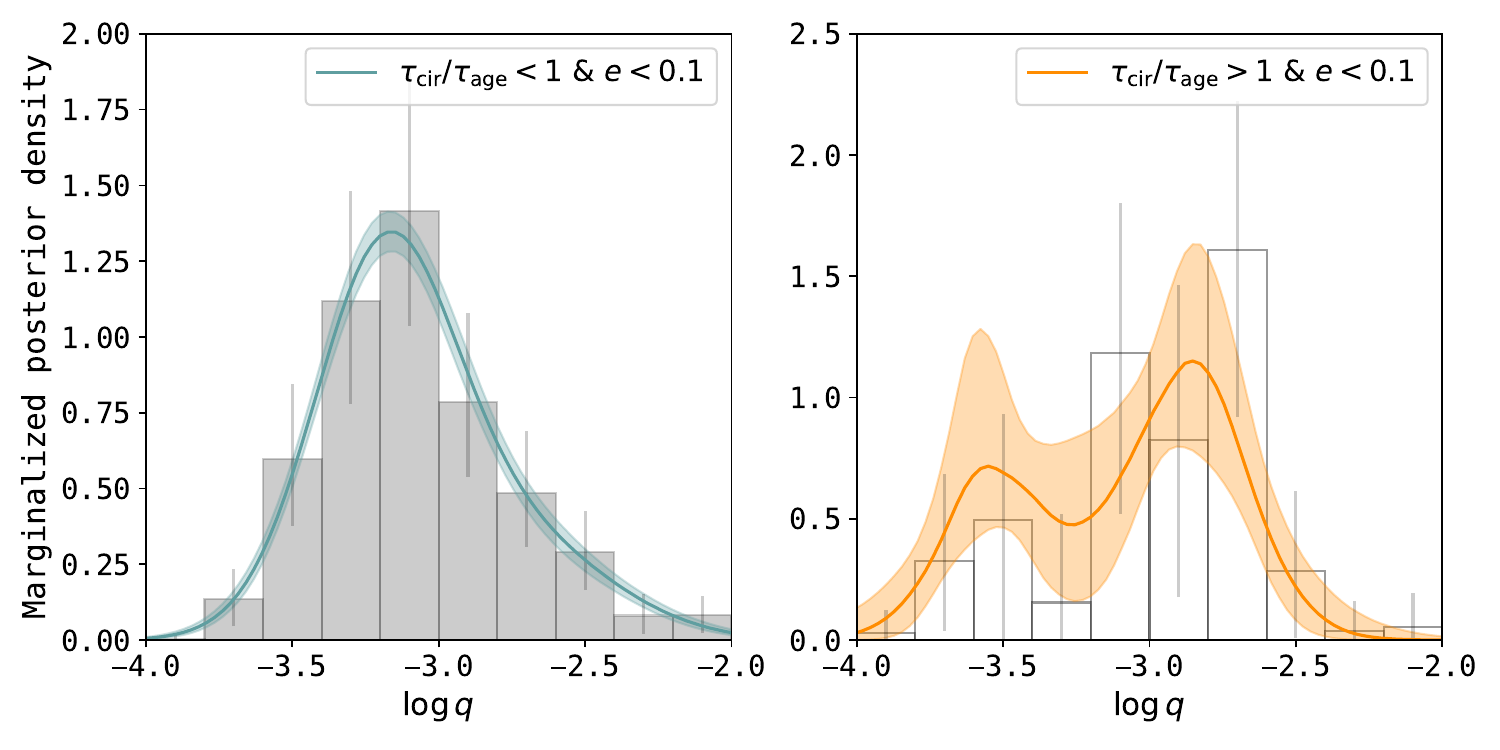}
  \caption{The marginalized posterior density $\log q$ for dynamically old and young systems on circular orbits.}
  \label{fig:histogram_model}
\end{figure*}

To investigate this potential trend, we perform a hierarchical Bayesian analysis to infer the mass ratio distribution while accounting for the uncertainty in $\tau_\mathrm{cir}/\tau_\mathrm{age}$. The methodology and detailed setup of the experiment is provided in Appendix  \ref{hierachical_bayesian_analysis}. We find that the Gaussian mixture modeling of mass ratio clearly identifies a bimodal structure in the disk migration candidates. We present the inferred distribution in Figure \ref{fig:histogram_model}. 

Since the sample sizes of both dynamically young samples and microlensing planets are still small, the significance of the dip in either case will need further validation with future observations. For example, the dip was not observed in previous microlensing study \citep{Suzuki_2016}, although the sample size in \cite{Suzuki_2016} ($N=30$) was about half of \cite{Zang_2025} ($N=63$). We should also note that host stars of microlensing planets have lower median mass ($\sim0.6M_\odot$) than our samples ($\sim1.1M_\odot$). So even though the dip location aligns in mass ratio, dip location in terms of planet mass would be different. In microlensing samples, the dip corresponds to $0.06M_\mathrm{J} < M_\mathrm{p}<0.3M_\mathrm{J}$, and in our samples the dip corresponds to $0.3M_\mathrm{J}<M_\mathrm{p}<1M_\mathrm{J}$. On a similar note, a dip in planet mass was also not recovered by the analysis of radial velocity sample in \cite{Bennett_2021}.

One optimistic interpretation is that the distribution of mass ratio and not planet mass reflects astrophysical effects. The lack of planets with specific mass ratio in both microlensing ($P\sim1-50 $ years) and disk migration ($P\sim 4-365$ days) samples, as opposed to the ubiquity in dynamically old samples ($P\sim1-7$ days) would be best explained if disk migration preferably brought the planets within this specific mass ratio range to the disk inner edge. Since disk migration candidates identified in this work all have relatively long orbital periods that are probably larger than periods expected from their respective disk truncation radii, it also suggests that disk migration for planets in other mass ratio regime can halt before reaching the inner edge.

Such selective disk migration mechanism with specific mass ratio is qualitatively consistent with the predictions of Type~III migration, while planets outside the dip region may be explained by Type~I and Type~II migration. In Type III migration, intermediate-mass planets—those massive enough to partially open a gap but not sufficient to clear the co-orbital region—can undergo rapid, runaway migration driven by torques from residual disk material \citep{Masset_2003,Papaloizou_2007}. The tendency for planets to open gaps in protoplanetary disks scales with their mass ratio \citep{Bryden_1999, Crida_2006}. The observed dip is roughly consistent with the mass ratio threshold for gap opening inferred from observations \citep{Kanagawa_2016}. Runaway migration has historically been proposed as one explanation for the relative paucity of massive hot Jupiters compared to their colder counterparts \citep{Udry_2003}, but has been difficult to further investigate observationally. Therefore, if significance of the dip in mass ratio increases with future observations, it presents a unique opportunity to probe the migration of giant planets.

Alternatively, a dip in the planet mass distribution has long been predicted by core accretion theory, where runaway gas accretion bifurcates the planet population, resulting in a dearth of planets with mass in the range \(10M_\oplus\leq M_\mathrm{p} \leq 100M_\oplus\) \citep{Ida_Lin_2004} especially around $0.2\mathrm{AU} < a< 3\mathrm{AU}$. Indeed, literature including \cite{Zang_2025} interprets the presence or absence of this dip as evidence for or against runaway accretion. With runaway accretion, however, it is difficult reconcile the difference in the mass ratio distribution between dynamically old and young Jupiters. Meanwhile, selective migration can explain the discrepancy more naturally.

\subsection{Prospects for future observations}
To further investigate the three trends identified in this work, it is first necessary to increase the sample of dynamically young hot Jupiters. For this, the validation and confirmation of planet candidates on wide orbits (longer expected $\tau_\mathrm{cir}$) plays a crucial role. Additionally, refining the eccentricity constraints for systems listed in Table~\ref{ident-dm} is important, as many currently have only upper limits or assume circular orbits.

Secondly, obliquity measurements of disk migration candidates will probe the rate and the extent of primordial misalignment discussed in Section \ref{sec:primordial_alignment}. Such kinds of observations will be complementary to recent obliquity surveys of warm Jupiters \citep[e.g.][]{Rice_2022}.

Lastly, it will be valuable to revisit the archival photometry with Kepler and TESS to search for nearby companions in the disk migration candidates. We recovered no extra transit signals in the candidate systems with a preliminary search using \texttt{opents}\footnote{\url{https://github.com/hpparvi/opents}}, an open source tool to search for transit signals in Kepler and TESS light curves. However, TTVs can smear out potential transit signals, and there is room for more tailored and detailed detrending of the light curves. On a similar note, a uniform search for nearby companions using homogeneous dataset is crucial for a quantitative analysis of disk migration fraction \citep[e.g.][]{Sha_2024}, which we discussed in Section \ref{discussion_nearby_comp}.  

Finally, we marked the 17 disk migration candidate hot Jupiters that are also Ariel Tier 3 targets with asterisks in Table~\ref{ident-dm}. The transmission and/or emission spectra obtained with Ariel will provide exciting opportunities to probe the elemental abundances of these planets including C/O, N/O and C/N \citep{Tinetti+2022}. For disk migration candidates, where carbon and oxygen abundances are expected to be dominated by the accretion of solids, a general expectation is a sub-stellar C/O ratio \citep{Oberg+2011,Madhusudhan+2016}.

\section{Conclusion}
In this paper, we identified close-in Jupiters that likely arrived via disk migration by leveraging on the idea that when circularization timescale of a planet is longer than system age ($\tau_\mathrm{cir}>\tau_\mathrm{age}$), HEM would not be able to complete in time. To do this, we empirically calibrated the tidal quality factor $Q_\mathrm{p}$ using the eccentricity distribution of 578 Jovian mass planets with measured masses and radii. We remind that the method used in this study encapsulates only the dissipation of equilibrium tides on a tidally locked planet. The samples are also heterogeneous in terms of discovery and analysis methods. Nevertheless, the typical value of $Q_\mathrm{p}$ derived in this work is consistent with that of Solar system Jupiter, and seems to be relatively indifferent of planet mass.

We find the following three trends in disk migration candidates: 

\begin{enumerate}
    \item \textbf{Low stellar obliquity:} For planets on circular orbits, we observe a cutoff in obliquity around $\tau_\mathrm{cir}\sim\tau_\mathrm{age}$. The fact that disk migration candidates are mostly on aligned orbits supports the notion that planets generally form in primordially aligned protoplanetary disks. This is consistent with the findings by previous works focusing on obliquity measurements of warm Jupiters and multiplanetary systems \citep{Rice_2022,Razdom_2024,Razdom_2025}. Among the disk migration candidates, two rare systems with non-zero obliquity have distant stellar companions, which could be responsible for primordial disk misalignment. As we began our experiment agnostic about the obliquity of planets in the sample, this cutoff also supports the validity of dynamical age as a proxy for disk migration. 

    \item \textbf{Preference for nearby companions:}
    We observe that dynamically young hot Jupiters on circular orbits are more likely to host nearby companions than the overall population of hot Jupiters. This is another support for dynamical age as proxy for disk migration, as HEM would likely disrupt nearby companions. On the other hand, most hot Jupiters are dynamically old and only a fraction of them have nearby companions. This is only expected if HEM is the dominant formation channel, or that disk migration also disrupts nearby companions more effectively at close-in orbits. However, rate of disruption in the case of disk migration should also not be too high, otherwise we need to assume that almost all hot Jupiters arrived via disk migration. This fails to explain the many planets on high obliquity.  

    \item \textbf{Dip in mass ratio:}
    There is a potential dip in the mass ratio of disk migration candidates, as opposed to the ubiquity of dynamically old hot Jupiters around $\log q\sim -3.2$. A dip in the same location was recently suggested for planets on much wider orbits \citep{Zang_2025}. A natural explanation for such a trend is that disk migration happens more efficiently for the planets in the dip, favoring them to reach the disk edge. Qualitatively, this is consistent with the prediction of Type~III (runaway) migration. The significance of the dip in both disk migration candidates and in microlensing samples needs to be tested with future observations. 

\end{enumerate}

Further investigation of these trends, as well as future characterization of disk migration candidates will provide important clues to the migration mechanism of hot Jupiters.

\section{Acknowledgments}
We thank the anonymous referee for insightful comments that helped improve the quality of this manuscript. We thank Judith Korth and Hannu Parvieinen for introducing \texttt{opents}, Jim Fuller, Teruyuki Hirano, Yasunori Hori, Yuji Matsumoto and Shigeru Ida for discussions on tidal efficiency, John Livingston for discussions on analysis methods and Shota Miyazaki for discussion on hierarchical Bayesian analysis. YK supported by JST SPRING, Grant Number JPMJSP2108 and JSPS Grant-in-Aid for JSPS Fellows Grant Number JP25KJ1036. This work is partly supported by JSPS KAKENHI Grant Number JP24H00017, JP24K00689, JSPS Bilateral Program Number JPJSBP120249910 and JSPS Grant-in-Aid for JSPS Fellows Grant Number JP25KJ0091. This research has made use of the NASA Exoplanet Archive \citep{Confirmed_planets_table}, which is operated by the California Institute of Technology, under contract with the National Aeronautics and Space Administration under the Exoplanet Exploration Program.

\bibliography{citation}{}
\bibliographystyle{aasjournalv7}

\appendix
\section{Hierarchical Bayesian Modeling of Mass Ratio}
\label{hierachical_bayesian_analysis}

\begin{table*}[ht]
\centering
\caption{Prior and posterior summaries for population hyperparameters}
\begin{tabular}{llccc}
\hline
\textbf{Parameter} & \textbf{Description} & \textbf{Prior} & \multicolumn{2}{c}{\textbf{Posterior (Mean ± 1$\sigma$)}} \\
\cline{4-5}
 & & & $k=1$ & $k=2$ \\
\hline
$\alpha_k$      & Log density in $k$-th bin of $\tau_\mathrm{cir}/\tau_\mathrm{age}$ & $\mathcal{N}(0, 5)$         & $6.24\pm3.65$ &  $-6.08\pm3.65$\\
$w_{k,1}$       & Mixture weight of first component         & $\mathrm{Dirichlet}(1, 1)$ & $0.69\pm0.14$ & $0.44\pm0.20$ \\
$w_{k,2}$       & Mixture weight of second component        & --                         & $0.31\pm0.14$ & $0.56\pm0.20$ \\
$\mu_k$         & Mean of first Gaussian component          & $\mathcal{N}(-3, 1.0)$         & $-3.19\pm0.03$ & $-3.50\pm0.18$ \\
$\delta\mu_k$   & Offset of second Gaussian mean            & $\mathrm{HalfNormal}(1)$    & $0.43\pm0.13$ & $0.63\pm0.21$ \\
$\sigma_{k,1}$  & Std. dev. of first component              & $\mathrm{HalfNormal}(1)$    & $0.24\pm0.02$ & $0.22\pm0.15$ \\
$\sigma_{k,2}$  & Std. dev. of second component             & $\mathrm{HalfNormal}(1)$    & $0.33\pm0.05$ & $0.25\pm0.13$ \\
$\gamma_{k,l}$  & Log density in $l$-th bin of $\log q$     & $\mathcal{N}(0, 5)$         & \multicolumn{2}{c}{See Figure~\ref{fig:histogram_model}} \\
\hline
\end{tabular}
\label{tab:hyperparams}
\end{table*}

To investigate the potential dip in the mass ratio distribution, we perform a hierarchical Bayesian analysis to model the mass ratio distribution of dynamically young and old planetary systems, while properly accounting for the uncertainty in $\tau_\mathrm{cir}/\tau_\mathrm{age}$. This requires simultaneously modeling the distribution of $\tau_\mathrm{cir}/\tau_\mathrm{age}$ and the logarithmic planet-to-star mass ratio. We adapt the methodology devised in \cite{Miyazaki_Masuda_2023}, who jointly modeled the distribution of stellar parameters and the occurrence rate of hot Jupiters in the CKS sample.

We first model the distribution of $\tau_\mathrm{cir}/\tau_\mathrm{age}$ using a step-function (histogram) prior:
\begin{equation}
p(\tau \mid \boldsymbol{\alpha}) = 
\sum_{k=1}^{K} \exp(\alpha_k) \, \mathbf{1}_k(\tau_\mathrm{cir}/\tau_\mathrm{age}),
\end{equation}
\noindent where the domain of $\tau_\mathrm{cir}/\tau_\mathrm{age}$ is divided into $K = 2$ bins corresponding to dynamically young ($\tau_\mathrm{cir}/\tau_\mathrm{age} > 1$) and dynamically old ($\tau_\mathrm{cir}/\tau_\mathrm{age} < 1$) regimes. The indicator function $\mathbf{1}_k(\tau_\mathrm{cir}/\tau_\mathrm{age})$ returns 1 if the value falls in the $k$-th bin and 0 otherwise. The height of each bin is given by $\exp(\alpha_k)$, where $\alpha_k$ is an unconstrained log-density parameter. The distribution is normalized such that $\sum_{k} \exp(\alpha_k) \, \Delta_k = 1$, where $\Delta_k$ is the width of the $k$-th bin.

Within each $k$-th bin, we test two models for the distribution of $\log q$. The first is a histogram model with:
\begin{equation}
p(\log q \mid \boldsymbol{\theta}, k) = \sum_{l=1}^{L} \exp(\gamma_{k,l}) \, \mathbf{1}_l(\log q),
\end{equation}

\noindent where \(\gamma_{k,l}\) is the log-density in bin \(j\) of $\log q$ in the $k$th-bin, and $\Delta_l$ is the bin width. These are normalized such that \(\sum_l \exp(\gamma_{k,l}) \, \Delta_l = 1\). This model provides a first look into the shape of the distribution. The other is a two-component Gaussian mixture model with:
\begin{equation}
\begin{split}
p(\log q \mid \boldsymbol{\theta}, k) &=
w_{k,1} \, \mathcal{N}(\log q \mid \mu_k, \sigma_{k,1})\\ 
&+w_{k,2} \, \mathcal{N}(\log q \mid \mu_k + \delta\mu_k, \sigma_{k,2}),
\end{split}
\end{equation}
where the means are ordered via $\mu_{k,1} = \mu_k$ and $\mu_{k,2} = \mu_k + \delta\mu_k$ to avoid label switching, and the weights satisfy $w_{k,1} + w_{k,2} = 1$. This model can capture potential dips or bimodal structure in the distribution, while remaining consistent with a unimodal shape when appropriate, as one of the mixture weights can shrink toward zero.

To determine the joint PDF given data $D$, $p(\boldsymbol{\alpha,\theta}|D)$, we evaluate the below log likelihood approximated using importance sampling \citep{Hogg_2010}:
\begin{equation}
\begin{split}
\ln p(D \mid \boldsymbol{\theta}, \boldsymbol{\alpha}) \approx
\sum_{j=1}^{N} \ln \left[ \frac{1}{S} \sum_{i=1}^{S}
\frac{
p(\log q_j^{(i)} \mid \boldsymbol{\theta}, k_j^{(i)}) \cdot 
\exp(\alpha_{k_j^{(i)}})
}
{
p_0(\log q_j^{(i)}, \tau_{\mathrm{cir},j}^{(i)}/\tau_{\mathrm{age},j}^{(i)})
} \right],
\end{split}
\end{equation}

\noindent where $\log q_j^{(i)}\sim p_0(\log q \mid D^j)$ is the $i$-th sample of log$q$ of the $j$-th planet resampled with the same method outlined in Section~\ref{sec:methods}. Accordingly, \(k_j^{(i)}\) denotes the index of the \(\tau_\mathrm{cir}/\tau_\mathrm{age}\) bin associated with the $i$-th sample of the $j$-th planet. Although the prior $p_0$ appears in the denominator, its absolute normalization does not affect the inference. 

We implement both these models using \texttt{PyMC} \citep{AbrilPla_2023,pymc_pipeline}, and the posteriors are sampled using no U-turn sampler, a gradient based flavor of the MCMC sampling algorithm. We assure convergence by checking that  $\hat{R}=1$. We summarize the inferred hyperparameters in Table~\ref{tab:hyperparams} and illustrate the resulting posterior distributions in Figure~\ref{fig:histogram_model}.



\end{document}